\documentstyle[12pt]{article}
\newcommand {\beq}{\begin{equation}}
\newcommand {\eeq}{\end{equation}}
\newcommand {\beqa}{\begin{eqnarray}}
\newcommand {\eeqa}{\end{eqnarray}}
\newcommand {\beqan}{\begin{eqnarray*}}
\newcommand {\eeqan}{\end{eqnarray*}}
\newcommand {\n}{\nonumber \\}

\newcommand {\eq}[1]{eq.~(\ref{#1})}

\newcommand {\Romannumeral}[1]{\uppercase\expandafter{\romannumeral#1}}

\newcommand {\del}{\partial}

\def\rref#1{(\ref{#1})}

\begin{document}
\setlength{\oddsidemargin}{0cm}
\setlength{\baselineskip}{7mm}

\begin{titlepage}
 \renewcommand{\thefootnote}{\fnsymbol{footnote}}
    \begin{normalsize}
     \begin{flushright}
																					KEK-TH-453\\
                     TIT-HEP-309\\
																					YITP-95-9\\
                    November 1995~~
     \end{flushright}
    \end{normalsize}
    \begin{Large}
       \begin{center}
         {\LARGE Renormalizability of Quantum Gravity\\
                 near Two Dimensions} \\
       \end{center}
    \end{Large}

  \vspace{5mm}

\begin{center}
           Hikaru K{\sc awai}\footnote
           {E-mail address : kawai@theory.kek.jp}\\
        {\it National Laboratory for High Energy Physics (KEK),}\\
		 {\it Tsukuba, Ibaraki 305, Japan,} \\
           Yoshihisa K{\sc itazawa}\footnote
           {E-mail address : kitazawa@phys.titech.ac.jp}\\
        {\it Department of Physics,
Tokyo Institute of Technology,} \\
            {\it Oh-okayama, Meguro-ku, Tokyo 152, Japan}\\
				{\it and}\\
		Masao N{\sc inomiya}\footnote
           {E-mail address : ninomiya@yukawa.kyoto-u.ac.jp}\\
        {\it Yukawa Institute for Theoretical Physics,}\\

{\it Kyoto University, Kitashirakawa sakyo-ku,
		Kyoto 606-01, Japan} \\

\vspace{5mm}

\end{center}
\begin{abstract}
\noindent
We study the renormalizability of quantum gravity
near two dimensions.
Our formalism starts with
the tree action which is invariant under the
volume preserving diffeomorphism.
We identify the BRS invariance which originates from the
full diffeomorphism invariance.
We study the Ward-Takahashi identities to determine the
general structure of the counter terms.
We prove to all orders that the counter terms can be supplied by the
coupling and the wave function renormalization of the tree action.
The bare action can be constructed to be the Einstein action form
which ensures the full diffeomorphism invariance.

\end{abstract}
\end{titlepage}
\vfil\eject

\section{Introduction}

As it is well known, the renormalizable field theories
are those with dimensionless coupling constants.
Those theories are also classically conformally
invariant. As a nonperturbative definition
of the continuum field theory, we may consider the
Euclidean statistical systems on the lattice.
The continuum field theory is defined in the
vicinity of the critical point of the system
where it exhibits the conformal invariance.

It is therefore reasonable to postulate that the short
distance limit of the consistent field theories
exhibits the conformal invariance.
This postulate holds nonperturbatively
and the nonlinear sigma models between two and four
dimensions are such examples.
Nevertheless we can devise a perturbative expansion
around two dimensions by studying these theories
in the $2+\epsilon$ dimensions.
In this expansion, $\epsilon$
is regarded to be a small expansion parameter
and the theory is weakly coupled at the short distance
fixed point of the renormalization group
if $\epsilon$ is small.

The Einstein gravity may fall into this category.
It is classically conformally invariant and renormalizable
in two dimensions. Furthermore it is topologically
invariant in two dimensions.
The Newton (gravitational) coupling constant is found
to be asymptotically free\cite{2+epsilon,KN}.
The investigations in the context of the
string theory and the matrix models have vindicated
the asymptotic freedom of the gravitational coupling constant
in two dimensions.
Therefore it is similar to the nonlinear sigma models
and it is tempting to conjecture that the short distance
fixed point in the renormalization group exists
beyond two dimensions. It implies the existence
of the consistent quantum gravity beyond two
dimensions.

In our investigation of this problem, the dynamics of the
conformal mode of the metric is found to be very different
from the rest of the degrees of freedom\cite{KKN1,KKN2,AKKN}.
Therefore we need to treat
these two different variables differently.
For this purpose,
we decompose the metric into the conformal factor
and the rest as $g_{\mu\nu} = \hat{g}_{\mu\rho}
(e^h)^{\rho}_{~\nu}e^{-\phi} = \tilde{g}_{\mu\nu}e^{-\phi}$.
Here we have also introduced a background metric
$\hat{g}_{\mu\nu}$.
The tensor indices of the fields are raised and lowered by
the background metric. $h_{\mu\nu}$ is a traceless symmetric
tensor.
The pure Einstein action in this parametrization is:
\beq
I_{Einstein} = \int {\mu ^{\epsilon}\over G}  e^{-{\epsilon\over 2}\phi}
\{\tilde{R}
- {\epsilon (D-1)\over 4} \tilde{g}^{\mu\nu}
\del _{\mu} \phi \del _{\nu} \phi \} ,
\label{origac}
\eeq
where $\int = \int d^Dx\sqrt{\hat{g}}$ denotes the
integration over the $D$ dimensional spacetime.
$\tilde{R}$ is the scalar curvature made out
of $\tilde{g}_{\mu\nu}$.
$G$ is the gravitational coupling constant
and $\mu$ is the renormalization scale to define it.

The $\beta$ function of the gravitational coupling
constant ($\beta _G$) in $D=2+\epsilon$ dimensions at the one loop level
is
\beq
\mu{\del \over \del \mu} G
= {\epsilon G} - {25-c\over 24\pi}G^2,
\label{betag}
\eeq
where $c$ counts the matter contents.
It shows that the theory is well defined at short
distance as long as $c < 25$.
The short distance fixed point of the $\beta$
function is $G^*=24\pi\epsilon /(25-c)$.

The pure Einstein action can be rewritten in the
following way by the change of the variables
with respect to the conformal mode:
\beq
I_{gravity} = \int {\mu ^{\epsilon}\over G}
\{\tilde{R}(1+a\psi + \epsilon b \psi ^2)
- {1\over 2} \tilde{g}^{\mu\nu}
\del _{\mu} \psi \del _{\nu} \psi \} .
\label{renac}
\eeq
In this expression, the kinetic term for the conformal
mode becomes canonical.
Classically $a^2 = 4 \epsilon b = \epsilon / 2 (D-1)$.

$a^2$ is an indicator of the conformal mode
dependence of the theory (conformal anomaly).
As it is well known the conformal anomaly is
synonymous to the $\beta$ functions.
Therefore the nontrivial $\beta$ function \rref{betag}
implies that
$a^2$ will be renormalized at the quantum level.
It will be shown that it is related to the $\beta$ function of $G$
as $a^2 G = \beta _G / 2(D-1)$ in section 4.
Therefore $a^2$ can be expanded by $G$ as
\beq
a^2 = {1\over 2(D-1)} (\epsilon -AG - 2BG^2 \ldots ).
\label{asq}
\eeq
We encounter $1/\epsilon$ poles if we expand
$a$ in terms of $G$ since $a^2$ starts with $\epsilon$.
For this reason we treat $a$ as another coupling
constant of the theory whose square is related to
$\beta _G$ by \rref{asq}.
We start with the tree action which generalizes
the Einstein action in this way.
We refer the generalized Einstein action in \eq{renac}
as the gravity action in this paper.

It is straightforward to couple the matter
in the conformally invariant way by adding the
following matter action.
\beq
I_{matter} =
\int
\{
 {1\over 2} \tilde{g}^{\mu\nu}
\del _{\mu} \varphi_i \del _{\nu} \varphi _i
-\epsilon b \tilde{R} \varphi ^2_i \},
\label{matter}
\eeq
where $i$ which runs up to $c$ counts the matter contents.
In this parametrization, we have rescaled the matter field
by the conformal mode to show the decoupling of the conformal
mode explicitly.

Although $G$ governs the dynamics of $h_{\mu\nu}$
field, the dynamics of the conformal mode
is governed by the coefficient $a$
which has a singular expansion in $G$.
This singular expansion arises due to the presence of the kinematical
poles\cite{KN}.
This is the origin of the difficulty to carry out
the $2 + \epsilon$ dimensional expansion of
quantum gravity.
Our formulation resums the kinematical poles to
all orders in $G$\cite{KKN1}.
For this purpose, we have proposed that we should treat $G$ and $a$
as independent couplings.
We consider
a tree level action which is invariant under the
volume preserving diffeomorphism.
The general covariance can be recovered by further
imposing the conformal invariance on the theory
with respect to the background
metric\cite{KKN2,AKKN}.
This requirement certainly determines the relation
between $G$ and $a$ at the classical level.
At the quantum level, this relation receives
corrections
due to the $\beta$ function just like \eq{asq}.

Concerning this background independence requirement,
we realize that the Einstein action is certainly
a solution for it. It is in fact to be a unique one.
Therefore we conclude that the bare action which
is obtained by adding the counter terms to the tree
level action is the Einstein action.
This point has been found to be the case at
the one loop level calculation\cite{AKKN}.
In our formulation, the tree action captures the dominant
dynamics whose corrections are small in the perturbation theory.
The question now is whether we can renormalize
the theory in this scheme to all orders
in the perturbation theory.

In \cite{Kitazawa}, one of us has proposed such
a proof based on the Ward-Takahashi identity.
This identity follows from the gauge invariance
of the theory just like in generic gauge theories.
However in that work, the identity is assumed
to be valid up to only the required orders
in the perturbation theory.
In this work we require that the WT identity
to be the exact identity on the theory.
Such a requirement
is very restrictive.
It leaves us no choice except to conclude
that the bare action is the Einstein action.
However we still assume that the tree level
action is invariant under the volume preserving
diffeomorphism only.  This is because we need to
treat the dynamics of $h_{\mu\nu}$ and $\psi$
fields differently.
In this paper we construct a proof
which shows that the
theory is renormalizable to all orders
within this scheme.
We have thus further clarified the structure of the bare action
and how the general covariance is ensured in our
formalism.

We further study the renormalization
of the relevant operators such as the
cosmological constant operator.
This question is certainly crucial
for the theory to be physically meaningful.
We prove that the cosmological constant operator
is multiplicatively renormalizable.
Just like in two dimensions, the anomalous
dimensions are $O(1)$ in general and
which forces us particular considerations.
However the anomalous dimensions are calculable
near two dimensions by the saddle point method.
Although the cosmological constant operator
is multiplicatively renormalizable, it may be
useful to consider the renormalized cosmological
constant operator which incorporates the quantum
effect.
The functional form of the renormalized cosmological
constant operator is fixed by requiring that
the renormalized cosmological constant operator
is background independent.
We argue that the bare cosmological constant operator
which is obtained by adding the necessary counter terms
to the renormalized cosmological constant operator
is again of the generally covariant form.

The organization of this paper is as follows.
In section two,
we set up the BRS formalism
and derive the exact Ward-Takahashi identities.
In section three, we solve the WT identity to determine
the bare action. We give the inductive proof of the
renormalizability in section four.
We show that the
divergences of the theory can be canceled by the counter terms
which can be supplied by the coupling constant
and the wave function renormalization of the tree action.
We also show that the bare action can be constructed to be
the Einstein action form.
In section five, we study the renormalization
of the cosmological constant operator.
We discuss a physical definition of the $\beta$ function
in quantum gravity.
We conclude in section six with discussions.

\section{BRS Invariance and Ward-Takahashi Identity}

In this section,
we set up the BRS formalism in quantum gravity and derive
the exact Ward-Takahashi identities.
These identities are the consequence of the
general covariance.

We adopt the action \rref{renac}
as the tree level action.
The coefficient $a$ appears with the single power
of $\psi$. Therefore there is a parity invariance
under the simultaneous change of the signs of $a$ and
$\psi$.
We can classify the effective action into the even and
odd parity sectors. $\psi$ field appears as the even and odd
powers in each sector.
Due to this parity invariance,
only $a^2$ appears in the quantum corrections
in the even parity sector.
In the odd parity sector, the situation is the same
apart from the over all factor of $a$.
Therefore these corrections can be expanded in terms
of $G$ by using the relation of \rref{asq}.
These arguments hold as long as we choose the
gauge fixing terms which also respect the parity
invariance.

The crucial symmetry of the theory is the invariance
under the diffeomorphism.
The metric changes under the general coordinate
transformation as:
\beqa
\delta g_{\mu\nu} & = &
\del _{\mu} \epsilon ^{\rho} g _{\rho\nu}
+{g}_{\mu\rho} \del _{\nu}\epsilon ^{\rho}
+ \epsilon ^{\rho} \del _{\rho} {g} _{\mu\nu} .
\eeqa
We have decomposed the metric into the conformal mode
and the rest as
$g_{\mu\nu} = \tilde{g}_{\mu\nu} \psi ^{4\over \epsilon}$.
The over all scale of $\psi$ is irrelevant in this paragraph.
Here $det \tilde{g} = det \hat{g}$ since $\tilde{g}=\hat{g}e^h$ .
In this decomposition, the general
coordinate transformation takes the
following form:
\beqa
\delta \tilde{g}_{\mu\nu} & = &
\del _{\mu} \epsilon ^{\rho} \tilde{g} _{\rho\nu}
+\tilde{g}_{\mu\rho} \del _{\nu}\epsilon ^{\rho}
+ \epsilon ^{\rho} \del _{\rho} \tilde{g} _{\mu\nu}
-{2\over D} \nabla _{\rho} \epsilon ^{\rho}
\tilde{g} _{\mu\nu} ,\nonumber \\
\delta \psi & = & \epsilon ^{\rho} \del _{\rho}
\psi + ((D-1) a+{\epsilon\over 4} \psi)
{2\over D} \nabla _{\rho} \epsilon ^{\rho}, \n
\delta \varphi _i  & = & \epsilon ^{\rho} \del _{\rho}
\varphi _i + ({\epsilon \over 4} \varphi _i )
{2\over D} \nabla _{\rho} \epsilon ^{\rho} ,
\label{gaugetr}
\eeqa
where the covariant derivative is taken with respect
to the background metric.
An arbitrary constant $a$ can be introduced here by the
constant shift of $\psi$, although this shift
is singular in $\epsilon$.
$a$ may be viewed as the vacuum expectation value of $\psi$.
The matter fields transform as above since we have scaled the matter
fields by a single factor of $\psi$ in \eq{matter}.

As we have explained, the odd parity sector has the single
power of $a$ as the overall factor and the even powers of $a$ always
appear apart from this overall factor.
\rref{gaugetr} is consistent with such a structure since
the single power of $a$ appears when the powers of $\psi$
are reduced by one.
Therefore this invariance can be enforced
on the theory by using the relation \rref{asq}
without expanding $a$ by $G$.

In order to prove the renormalizability of the theory,
we set up the BRS formalism\cite{ZJ}.
The BRS transformation of these fields $\delta _B$
is defined by replacing the gauge parameter by the
ghost field
$\epsilon ^{\mu} \rightarrow C^{\mu}$.
The BRS transformation of $h_{\mu\nu}$ field
is defined through the relation $\tilde{g} =
\hat{g}e^h$.
The BRS transformation of ghost, antighost
and auxiliary field is
\beqa
\delta _B C^{\mu}& = & C^{\nu} \nabla _{\nu} C^{\mu},
\nonumber \\
\delta _B \bar{C}^{\mu} & = & \lambda ^{\mu},
\nonumber \\
\delta _B \lambda ^{\mu} & = & 0.
\label{brstr}
\eeqa
The BRS transformation can be shown to be
nilpotent $\delta _B^2 = 0$.

We denote $A_i =
(h_{\mu\nu} , \psi, \varphi _j)$. We also introduce a gauge fixing
function $F_{\alpha} (A_i)$.
It is an arbitrary function of $A_i$ with dimension one.
We assume that it respects the parity invariance
of the tree action.
The gauge fixed action is
\beq
S = I +
{\mu ^\epsilon \over G}
\int
[- \lambda _{\alpha}\lambda ^{\alpha}
+\lambda ^{\alpha} F_{\alpha} - \bar{C}^{\alpha}
\delta _B F_{\alpha} - K^i \delta _B A_i
-L_{\alpha} \delta _B C^{\alpha} ].
\label{gaufxac}
\eeq
Here we have introduced sources $K$ and $L$ for the
composite operators. $I=I_{gravity} + I_{matter}$
is the total action without the BRS exact terms.
In what follows, $I$ will be referred as
the tree action.

The partition function is
\beq
Z=e^W=\int [dAdCd\bar{C}d\lambda]
exp(-S+{\mu^{\epsilon}\over G}
\int[J^iA_i+\bar{\eta}_{\alpha}C^{\alpha}
+\bar{C}^{\alpha}\eta_{\alpha} + l_{\alpha}\lambda^{\alpha}]).
\eeq
By the change of the variables with the BRS transformation form,
we obtain the Ward-Takahashi identity for the generating
functional of the connected Green's functions:
\beq
\int (J^i{\delta \over \delta K^i} -
\bar{\eta}_{\alpha}{\delta \over \delta L_{\alpha}}
+\eta_{\alpha}{\delta \over \delta l_{\alpha}})W
= 0 .
\label{wtw}
\eeq

The WT identity for the effective action is
obtained by the Legendre transformation:
\beq
\int[
{\delta \Gamma \over \delta A_i}
{\delta \Gamma \over \delta K^i} +
{\delta \Gamma \over \delta C^{\alpha}}
{\delta \Gamma \over \delta L_{\alpha}}
-{\mu^{\epsilon}\over G}\lambda ^{\alpha}
{\delta \Gamma \over \delta \bar{C}^{\alpha}}
]
= 0 .
\label{wtpv}
\eeq
In order to make the above expression finite,
we need to add all necessary counter terms to
$S$.
The bare action $S^0$ obtained in this way
satisfies the same equation:
\beq
\int [
{\delta S^0 \over \delta A_i}
{\delta S^0 \over \delta K^i} +
{\delta S^0 \over \delta C^{\alpha}}
{\delta S^0 \over \delta L_{\alpha}}
-{\mu^{\epsilon}\over G}\lambda ^{\alpha}
{\delta S^0 \over \delta \bar{C}^{\alpha}}]
= 0 .
\label{wts}
\eeq
On the other hand, \eq{wtw} follows from \eq{wts}
in dimensional regularization.
To simplify notations, we introduce an auxiliary field $M_{\alpha}$
and add to the action the combination $-{\mu^{\epsilon}\over G}
\int M_{\alpha}\lambda^{\alpha}$
in such a way that ${\mu^{\epsilon}\over G}
\lambda ^{\alpha}=-{\delta \Gamma\over \delta
M_{\alpha}} = -{\delta S\over \delta M_{\alpha}}$.
Then the left hand side of \eq{wtpv} and \eq{wts} become
homogeneous quadratic equations which we write symbolically
as $\Gamma *\Gamma$ and $S^0*S^0$ .

In our derivation of the WT identities, we have assumed the
invariance of the bare action under the gauge transformation
\rref{gaugetr}. However we start with the tree level
action which possesses only the volume preserving
diffeomorphism invariance.  The crucial question is
whether we can choose the counter terms of the
theory in such a way to satisfy these identities
by starting with such a tree action. We answer affirmatively
to this question in this paper.

\section{Analysis of the Bare Action}

In this section, we solve \eq{wts} to determine $S^0$.
$S^0$ will be simply denoted by $S$ in this section.
Let us examine the general structure of the bare action.
By power counting, it has to be a
local functional of fields and sources
with the dimension $D$.
We also have the ghost number conservation rule
and its ghost number has to be zero.
By these dimension and ghost number considerations,
it is easy to see that
$K$ and $L$ appear only linearly in $S$:
\beq
S = \int {1\over G^0}
[ - K^i(\delta_B' A_i) -L_{\alpha}(\delta_B' C^{\alpha})]
+\tilde{S},
\label{kl}
\eeq
where $\delta_B'$ denotes most general BRS like
transformations
with the correct dimension and ghost number.
It is also easy to see that there are no $\lambda$ and hence
no $\bar{C}$ dependence in $\delta_B'$.
Since $\lambda$ has dimension $1$, $\tilde{S}$ can be
at most quadratic in $\lambda$:
\beq
\tilde{S} = \int {1\over G^0}[
-{1\over 2} \tilde{E}_{\alpha\beta}\lambda^{\alpha}\lambda^{\beta}
+\lambda^{\alpha}\tilde{F}_{\alpha}
+\tilde{L}],
\label{gaugefx}
\eeq
where $\tilde{E}_{\alpha\beta}$ and $\tilde{F}_{\alpha}$ are
general local functions of $A,C$ and $\bar{C}$
with dimension zero and one respectively.
$G^0$ is the bare gravitational coupling constant
and it is the only quantity with dimension $-\epsilon$.

We denote below by $\theta^i$ the set of all anticommuting fields
$K^i,C^{\alpha},\bar{C}^{\alpha}$ and $x_i$ all
commuting fields $A_i,L_{\alpha},M_{\alpha}$.
The fundamental equation for the action $S$
takes then the form:
\beq
{\del S\over \del x_i}{\del S \over \del \theta _i} = 0.
\label{funeq}
\eeq
The equation \rref{funeq} is invariant under the following canonical
transformations.
Let us make the change of the variables
$(\theta, x) \rightarrow (\theta ' ,x')$:
\beqa
x_i &=& {\del \varphi \over \del \theta _i}(x',\theta ),\n
\theta '_i &=& {\del \varphi \over \del x'_i}(x',\theta ) .
\label{cantr}
\eeqa
We can verify that we recover the equation \rref{funeq}
in the new variables.
It has been shown that the set of the canonical transformations
\rref{cantr} is the most general set of the transformations
which leaves the equation \rref{funeq} invariant\cite{ZJ,BH}.

Let us write these transformations in the infinitesimal form:
\beqa
x_i'& = & x_i - {\del \varphi\over \del \theta^i},\n
(\theta^i)' & = & \theta^i + {\del\varphi\over \del x_i},
\eeqa
the action $S(\theta^i ,x_i)$ changes as
\beq
S(\theta',x') -S(\theta ,x) =  \Delta \varphi,
\eeq
where
\beq
\Delta = {\del S\over \del \theta _i}{\del\over \del x_i}
+{\del S\over \del x_i}{\del\over \del \theta _i}.
\eeq
We also have the following relation:
\beq
\Delta ^2 = [-{\del\over \del\theta^j}
(S*S)]
{\del\over \del x_j} +
[{\del\over \del x_j}
(S*S)]{\del\over\del\theta^j}.
\eeq
Therefore if $S$ is the solution of the equation \rref{funeq},
$S + \Delta \varphi$ is also the solution of it
since this is an infinitesimal canonical transformation
of the fields.
We call $\Delta \varphi$ a BRS exact solution of
the equation \rref{funeq}.

Since we are studying the Einstein gravity, the Einstein
action is the only generally covariant action with the
dimension $D$.
The action \rref{gaufxac} with the Einstein action for $I$
certainly satisfies the equation \rref{funeq}.
However the solution is not unique due to the
freedom in association with the canonical transformation
of the fields. On the other hand
it is the only freedom of the solutions of \rref{funeq}.
Therefore the equations of \rref{kl} and \rref{gaugefx}
have to be interpreted by the canonical transformations.
Physically the canonical transformations correspond to
the freedom in association with the wave function
renormalization and the gauge fixing procedure.

\section{Inductive Proof of the Renormalizability}

In this section, we construct an inductive proof of the
renormalizability of quantum gravity by the $2+\epsilon$
dimensional expansion approach.

Our analysis is based on the expansion of
the effective action by the gravitational coupling
constant $G$:
\beq
\Gamma = \sum _{l=0}^{\infty} \Gamma _{l},
\eeq
in which $\Gamma _0$ is the tree level action S.
We define $\Gamma _l$ to be the effective action
of $G^{l-1}$ order.
Our formalism contains two dimensionless parameters
$G$ and $a$.
Although
$a^2$ possesses the expansion by $G$ as in \rref{asq},
it starts with the quantity of $O(\epsilon )$ and the expansion of $a$ by $G$
is singular.
Therefore the effective
action can be expanded by $G$ apart from the overall
factor of $a$ in the odd parity sector.
$a$ is regarded as $(G)^0$ and the expansion
of the effective action by $G$ should be understood
in this sense.

Hence the effective action $\Gamma$ and the bare action
$S_0$ consist of the even and odd sectors as
\beqa
\Gamma &=& \Gamma _{even} + a \Gamma _{odd} ,\n
S^0 &=& S^0_{even} + a S^{0}_{odd},
\eeqa
where we have written $a$ dependence explicitly.
$\Gamma _{even(odd)}$ and $S^0_{even(odd)}$
can be expanded in $G$ alone by using
\eq{asq}.
What we would like to prove is that we can choose
$S^0$ which makes $\Gamma$ finite in such a way that
$\Gamma * \Gamma =0$. In dimensional regularization,
the bare action $S^0$ also satisfies $S^0 * S^0 = 0$.

In order to determine the effective action
at $G^{l-1}$ order, the $l$ loop level computation is required.
$a^2$ is also determined up to $G^l$ order by this computation.
$\Gamma _l$ differs from the conventional $l$ loop level effective
action since $a^2$ can be expanded in $G$.
Hence it also receives contributions from the
lower loop level.

We assume as an induction hypothesis that we have been able to
construct the bare action $S^0_{l-1}$
which satisfies $S^0_{l-1} * S^0_{l-1} = 0$
and renders $\Gamma $ finite
up to $G^{l-2}$ order by the $l-1$ loop level
computation.
Namely $\Gamma _{k}$ with $k \leq l-1$ are assumed
to be finite.
We denote $a$ which appears in $S^0_{l-1}$ as
$a_{l-1}$ and it is regarded as $(G)^0$.
$S^0_{l-1}$ consists of the even and the odd parity sectors
and the odd parity sector is multiplied by $a_{l-1}$.
The situation is the same with the effective action $\Gamma$
and the both $\Gamma _{even}$ and $\Gamma _{odd}$
are assumed to be finite up to $G^{l-2}$ order.
$a^2_{l-1}$ is assumed to be determined up to order $G^{l-1}$:
\beq
a^2_{l-1} = {1\over 2(D-1)} (\epsilon -AG  \ldots
- \lambda ^1_{l-1}G^{l-1}).
\eeq

Although the bare action is taken to satisfy
$S^0 * S^0 = 0$,
we have adopted the tree level action $S$ in
such a way that $S*S$
is of higher orders in $G$.
This choice is motivated by the presence of the
conformal anomaly in quantum gravity.
By starting with such a tree action,
our formalism can handle
the dynamics of the conformal mode
which is influenced by the conformal anomaly.
{}From our basic equation $\Gamma * \Gamma =0$,
we find the following relation at $G^{l-2}$ order:
\beq
S*\Gamma _l + \Gamma _l *S = \Delta \Gamma_l
= -\sum _{k=0}^{l-1}\sum _{m=0}^{k}
\Gamma _m * \Gamma _{k-m}.
\label{s*g}
\eeq
We recall that $\Gamma _l$ is $O(G^{l-1})$.
The right hand side of this equation
has to be at least $O(G^{l-2})$
by the inductive assumption
and we only consider
the quantities
of $G^{l-2}$ order in this equation.
By the induction hypothesis, the
right hand side of this equation is finite.
The reason is that it involves only $\Gamma _k$
with $k \leq l-1$.
If $a^2_{l-1}$ is obtained in this equation, we expand
it by $G$.
This expansion terminates at order $G^{l-1}$
by the inductive assumption.
Obviously we find no divergence
by doing that on the right hand side.
On the other hand, the left hand side
contains $\Gamma _l$ which is divergent
in general.
This equation therefore determines
the possible structure of the divergent part of
$\Gamma _l^{div}$ at $G^{l-1}$ order.

In the perturbative expansion of the field theory,
all divergences at the $l$ loop order are guaranteed to be local
as long as all subdiagrams are subtracted to be finite.
It is because such divergences can be made finite by
differentiating the external momenta.
We have assumed by the induction hypothesis that the effective
action has been made to be finite up to $G^{l-2}$ order.
At $G^{l-1}$ order, all subdiagrams are at most $G^{l-2}$
order.
We can then conclude that all divergences at
$G^{l-1}$ order are local by using the above argument.
Here we would like to discuss the treatment of the divergences
of $a^2/\epsilon$ type. In the leading order, they can be
regarded as finite. In the minimal subtraction scheme,
we need not subtract them. However we find divergences
if we expand them by $G$ to higher orders. We need to subtract them even
in the minimal subtraction scheme in higher orders.
Another possibility is to subtract them
from the leading order as a whole.  Such a subtraction scheme
might have some advantage in our formalism
since we can subtract the class of terms in consideration
at once.

We have thus reduced the question of the renormalizability  to
that of finding
the most general solutions of \eq{s*g}. The similar problems
have been investigated extensively in gauge theories\cite{ZJ,BH}.
We can find the following solutions of this equation
based on the results of such investigations.

Let $\varphi$ be a local functional of fields at
$G^l$ order. Then $\Delta \varphi$ is a local functional
of fields at $G^{l-1}$ order. The divergence
of this form is consistent with \eq{s*g} since
$\Delta ^2 \varphi = 0$ at $G^{l-2}$ order.
This type of the divergence is called as the
BRS exact part. $\Gamma _i^{div}$ can be decomposed
into the BRS exact part and the rest in general.
We call the rest of the divergence as
the nontrivial solution of \eq{s*g}.

The divergences of the tree action form
\eq{renac} and \eq{matter} which are of order $G^{l-1}$
do satisfy this equation.
It is because the tree action is generally
covariant to the leading order.
However
this equation allows more general
classes of the divergences which
can be seen as follows.
Let us consider a generic local action which is invariant
under the volume preserving diffeomorphism.
It is easy to see that such an action is invariant under the
gauge transformation \eq{gaugetr}
if it is invariant under
the following conformal transformation:
\beqa
\delta \psi &=& (D-1)(a+2\epsilon b\psi )\delta \bar{\phi} , \n
\delta \varphi _i &=& 2\epsilon b(D-1)\varphi _i \delta \bar{\phi} , \n
\delta \hat{g}_{\mu\nu} &=& - \hat{g}_{\mu\nu} \delta \bar{\phi} .
\label{conftr}
\eeqa

When we
plug such an action as $\Gamma _l$ on the left hand side of \eq{s*g},
we find that it is proportional to
the conformal anomaly of the action.
The conformal anomaly vanishes in the two dimensional
limit if the action becomes conformally invariant in two dimensions.
The simple pole divergences of this type thus result
in the finite conformal anomaly.
Therefore such divergences are
consistent with \eq{s*g}.

The nontrivial divergences can be classified
into the two types: those with the simple
pole in $\epsilon$ and those with higher
poles in $\epsilon$.
{}From the considerations we have just gone through,
we find that the higher pole divergences of
the tree action form are consistent with our basic
equation.
The simple pole divergences which are invariant under
the volume preserving diffeomorphism are
also consistent with \eq{s*g}
if they are conformally invariant in the two dimensional limit.
It is because $\Delta \Gamma _l^{div}$
is given by the finite conformal anomaly for such
divergences.

Through these considerations, we have found very general solutions
for the possible divergences which are consistent with \eq{s*g}.
We have classified them into the BRS exact and the nontrivial
solutions. The nontrivial solutions are classified into the
two different types. They are those of the tree action type
and those with the finite conformal anomaly type.
It is physically very plausible that
they are the only solutions of this equation.
As the major conjecture in this proof we assume
that the only solutions of \eq{s*g} are those
we have found in this section.

In this model, the only operator which is
invariant under the volume preserving diffeomorphism and
which is conformally
invariant in two dimensions is
$\int \tilde{R}$.
This operator is invariant under the
transformation \eq{gaugetr} modulo $O(\epsilon )$.
We adopt this operator and the gravity
and the matter actions as the independent operators.
$\psi$ field transforms in a specific way
in this gauge transformation. The conformal
transformation \eq{conftr} is the
specific type which is a part of the gauge transformation.
We remark that the operator
$\int \tilde{g}^{\mu\nu} \del _{\mu}\psi \del_{\nu}\psi$
is not invariant under \eq{conftr} in two dimensional limit
in our sense
since we regard $a$ as a finite coupling constant.
$\int \tilde{g}^{\mu\nu} \del _{\mu}\varphi _i \del_{\nu}\varphi _i$
will be regarded as the matter action to the leading order of
$\epsilon$.

Therefore at $l$ loop level, new divergences
of the following form may arise:
\beq
{\mu^{\epsilon}\over G}\int[
{\lambda ^1_l G^l\over \epsilon}\tilde{R}]
+ tree~ action .
\eeq
Here we denote the residue of the simple pole
at the $l$ loop level in association with $\int \tilde{R}$
by $\lambda ^1_l$.
The divergences of the gravity action form can be subtracted
by the renormalization of the gravitational coupling constant $G$.
The divergences of the matter action type can be subtracted
by the wave function renormalization of $\varphi _i$.
We assume here that the gauge fixing function $F_{\alpha}(A_i)$ does not
depend on the matter fields for simplicity.
In order to cancel the remaining divergence,
we need the counter term
$-\mu^{\epsilon}\int (\lambda ^1_lG^{l-1}/\epsilon )\tilde{R}$.
The bare action constructed in this way have to
be the Einstein action form in order to satisfy $S^0 * S^0 = 0$
as we have found in the previous section.
The pure Einstein action in the parametrization we have adopted is
\beq
{\mu^{\epsilon}\over G}\int [\tilde{R}({2(D-1)a^2\over\epsilon}+
a\psi +\epsilon b\psi ^2)
-{1\over 2}\del_{\mu}
\psi\del_{\nu}\psi\tilde{g}^{\mu\nu}].
\label{eineq}
\eeq
Therefore $G^l$ order part of $2(D-1) a^2$ is determined
from this requirement to be
$-\lambda ^1_l G^l$.
We point out that $\lambda ^1_l$ itself does not depend
on the $G^l$ order part of $2(D-1) a^2$ since it comes from the
quantum loop effect.

We still need to study the BRS exact divergences which
can be expressed as $\Delta \varphi$.
The general form of $\varphi$ is:
\beq
\varphi = \int
[K^i\Psi '_i +L_{\alpha}\Theta '^{\alpha}_{~\beta}C^{\beta} +
\bar{C}^{\alpha}
(F'_{\alpha} + \lambda^{\beta}E'_{\alpha\beta})],
\eeq
where $\Psi '$ and $\Theta '$ are general local functions of
$A,C,\bar{C}$ with dimension zero and vanishing ghost number.
As we have explained, the BRS exact part can be associated with
a canonical transformation on the fields.
Here we consider the physical implications of these
canonical transformations.
Under this transformation,
the part of $S^0$ linear in $K$ and $L$ changes as:
\beqa
K^i\delta_BA_i &\rightarrow& K^i\delta_BA_i
+K^i\delta_B(\Psi '_i)
-K^i{\del \delta_BA_i\over \del A_j}
\Psi '_j
+K^i{\del \delta_BA_i\over \del C^{\alpha}}
\Theta'^{\alpha}_{~\beta}C^{\beta},\n
L_{\alpha}\delta_BC^{\alpha} &\rightarrow&
L_{\alpha}\delta_BC^{\alpha} -
L_{\alpha}\delta_B(\Theta '^{\alpha}_{~\beta}C^{\beta})
+L_{\alpha}{\del \delta_BC^{\alpha}\over
\del C^{\beta}}\Theta '^{\beta}_{~\gamma}C^{\gamma}
-L_{\alpha}{\del \delta_BC^{\alpha}\over \del A_i}
\Psi '_i.
\eeqa

These infinitesimal deformations can be
interpreted as the change of the functional form of
the BRS transformation in association with
the wave function renormalization of the fields.
Note that
the functional form
of the BRS transformation has to change
in terms of the renormalized variables,
although
the functional form of the BRS transformation
remains the same in terms of the bare fields.
The renormalized BRS transformation continues to be nilpotent.
The rest of the BRS exact part causes the renormalization
of the gauge fixing part.

By defining the bare action at $l$ loop level
\beq
S^0_l=S^0_{l-1} - \Gamma _l^{div} +
higher~orders ,
\label{news0}
\eeq
it is possible to render $\Gamma$ finite up to order $G^{l-1}$.
The BRS exact counter terms can be interpreted as
the  renormalization of the wave functions
and the gauge fixing part.
The rest of the counter terms can be interpreted
as the coupling constant renormalization
of the tree level action.
The higher order terms in \eq{news0} has to be chosen
in such a way that
$S_l^0$ satisfies $S^0_l * S^0_l = 0$ exactly.
As we have explained in the previous section,
such a bare action has to be the gauge fixed
Einstein form modulo the canonical transformation
of the fields. Since $\Gamma _l^{div}$ is also of this type,
we can construct such a bare action by integrating
these infinitesimal deformations.
$a^2$ is now determined up to order $G^l$.

When we obtain $S^0_{l}$ from $S^0_{l-1}$,
we substitute $a_{l}$ for $a_{l-1}$.
The only difference which has been brought about
by this change is the addition of the
required counter term of
$\int \tilde{R}$ type at order $G^{l-1}$
apart from the change of  the definition of $a$
up to order $G^{l-1}$.
Now the circle is complete and we have proven
the renormalizability
of quantum gravity near two dimensions.
The major assumption we have made in this proof is that
the solutions we have found in this section
exhaust the solutions of \eq{s*g}.

Under this very plausible assumption,
we have established that this model is renormalizable
to all orders with the following bare action:
\beq
{\mu^{\epsilon}Z\over G}\int
[\tilde{R}(Z_G +
a\psi
+\epsilon b\psi ^2)
-{1\over 2}\del_{\mu}\psi\del_{\nu}\psi\tilde{g}^{\mu\nu}]
+\int
[
{1\over 2}\del_{\mu}\varphi _i\del_{\nu}\varphi _i\tilde{g}^{\mu\nu}
-\epsilon b\tilde{R}\varphi ^2_i] ,
\label{bareeg}
\eeq
where $Z_G = 1-\lambda ^1/ \epsilon$ and $2(D-1)a^2 =
\epsilon -\lambda ^1$.
Here we have omitted the BRS exact part of the action.
Of course we also need counter terms which correspond
to the wave function renormalization of the action.

Let us consider the physical significance of the
coefficient $a^2$. We note that it measures the
conformal anomaly of the theory.
At the critical point where $a$ vanishes, the
conformal mode becomes indistinguishable
from the scalar fields which couple to the
gravity in the conformally invariant way.
The $Z_2$ invariance under $\psi \rightarrow -\psi$
is restored at the critical point
since the odd parity sector of the effective action
vanishes. We have suggested that this $Z_2$ invariance
may distinguish the different phases of
quantum gravity\cite{AKKN}.

We would like to interpret this phenomenon as the
signature of the
conformal invariance.
$a^2$ can be expanded in $G$ as follows:
\beq
2(D-1) a^2 G = \epsilon G -AG^2 -2BG^3 \ldots
\label{betan}
\eeq
This quantity measures the conformal anomaly
of the theory as a function of $G$.
For the particular value of $G$, it vanishes
and the theory becomes conformally invariant
in the sense we have just explained.
As it is well known the $\beta$ function
is related to the conformal anomaly.
Therefore it is reasonable to adopt \rref{betan} as the $\beta$
function of $G$ by resorting to this connection.
We also recall that $G$ is the gravitational coupling constant.
It measures the strength of the coupling of $h_{\mu\nu}$
field at the momentum scale $\mu$.

In the conventional definition of the $\beta$
function, it is defined through the bare
gravitational coupling constant:
\beq
{1\over G^0} = {\mu ^{\epsilon}\over G} Z'_G .
\label{bareg}
\eeq
The $\beta$ function of $G$ is obtained by demanding
that the bare quantity is independent
of the renormalization scale $\mu$.
However there is an ambiguity in this procedure
since the bare gravitational coupling constant
changes if we rescale the conformal mode\cite{KN}.
The relation \rref{betan} is free from such
an ambiguity since this ambiguity does not alter the
classical relation.

Therefore we adopt the right hand side of \eq{betan}
as the $\beta$ function of $G$.
We expect that this $\beta$ function is also obtained
by the conventional procedure since $Z'_G=ZZ_G$
in our scheme.
Through the conventional procedure
we find that $\mu {\del\over \del \mu}G=\epsilon GZ'_G
/(1-G{\del\over \del G})Z'_G$.
If $Z=(1-G{\del\over \del G})Z'_G$,
we find the same $\beta$ function with \eq{betan}
by the conventional method.
In our renormalization procedure, we have classified the
nontrivial solutions of \eq{s*g} into those with the conformal anomaly
and those with the vanishing conformal anomaly.
The latter is associated with the higher poles in $\epsilon$
in general while the former is associated with the simple pole.
This classification must be generic in field theories.
We expect that such a classification of divergences underlies the pole
identities which ensure the finiteness of the conformal anomaly.

Before concluding this section, we evaluate the conformal anomaly
of the bare action with respect to the background metric.
Under the conformal transformation \eq{conftr}, the bare action
\eq{bareeg} changes as:
\beq
{\mu^{\epsilon}\over 2 G}\int (\epsilon Z_G -2(D-1)a^2)\tilde{R}_r\delta
\bar{\phi},
\eeq
where $\tilde{R}_r = Z\tilde{R}$ is the renormalized operator.
The coefficient of $\tilde{R}_r$ is called as a trace anomaly coefficient.
It is expressed in terms of the $\beta$ functions since $\beta _G =
\epsilon Z_G G$.
We observe that it certainly vanishes by the construction.
We have proposed to construct the quantum gravity by requiring that
it does not depend on the conformal mode of the background
metric\cite{KKN2,AKKN,Kitazawa}.
We have shown how this requirement is fulfilled in our renormalization
scheme.

\section{Renormalization of the cosmological constant operator}

In this section, we study the renormalization of the
cosmological constant operator. The renormalization
of the relevant spinless operators can be done
in the analogous procedures.
This problem has been studied extensively in our
previous works\cite{KKN1,KKN2,AKKN}.
We prove that the cosmological constant operator is
multiplicatively renormalizable by using the WT identity.
However the quantum corrections are $O(1)$
in general. It may be useful to define
the renormalized cosmological constant operator
which incorporates the large quantum renormalization
effect.
We can determine the functional form of the renormalized
cosmological constant operator by requiring the background
metric independence.
We further make the relation
between the bare and the renormalized cosmological constant
operators explicit in this section.

The bare cosmological constant operator
is of the generally covariant form:
\beqa
\int d^Dx \sqrt{g} & = &
\int
(1+{\epsilon b\over a}\psi )^{2D\over \epsilon} \n
& \sim &
\int exp({1\over a}\psi -{\epsilon \over 8a^2}\psi ^2
\ldots ) .
\label{barecos}
\eeqa
This operator is invariant under the gauge transformation \eq{gaugetr}
for an arbitrary value of $a$.
We consider the infinitesimal perturbation of the
theory by the cosmological constant operator.
The new tree action is $S + \Lambda \tilde{S}$ where
$\tilde{S}$ is the cosmological constant operator.
The effective action can also be expanded in $\Lambda$ as
$\Gamma +\Lambda \tilde{\Gamma}$.
The WT identity \eq{wtpv} becomes at $O(\Lambda)$ as:
\beq
\Gamma *\tilde{\Gamma} +\tilde{\Gamma}*\Gamma = 0 .
\eeq

The new tree action satisfies this WT identity.
We consider the perturbative expansion of the
theory by $G$.
The inverse powers of $a$ appear in the cosmological
constant operator.  We regard the inserve powers of $a$ as $O(1)$
in this paragraph.
Let us assume by the induction hypothesis that
we have made the effective action finite up to $G^{l-2}$
order.
The effective action at $G^{l-1}$ order satisfies the following
WT identity:
\beq
S*\tilde{\Gamma}_{l} + \tilde{\Gamma}_{l}*S
+\tilde{S}*\Gamma _{l} + \Gamma _{l}*\tilde{S}
 =  \sum _{k=1}^{l-1}(
\Gamma _k*\tilde{\Gamma}_{l-k}
+\tilde{\Gamma}_{l-k}*\Gamma _k ) ,
\eeq
where we have also expanded $\tilde{\Gamma}$ by $G$.
This equation can determine the general structure of the counter terms of
dimension zero which are required to cancel the
divergent part of $\tilde{\Gamma}_{l}$.

We find that the divergences of
the bare cosmological constant operator form is consistent
with this equation.
Unlike the dimension two operators, we cannot find other
nontrivial solutions.
We conjecture that the only BRS nontrivial
divergences of dimension zero take the bare cosmological
constant operator form.
Hence the cosmological constant operator is
multiplicatively renormalizable.
However the anomalous dimensions are not small even in the
perturbation theory since it is $O({G\over a^2})$ and
$a^2$ is $O(G)$ at short distance. Therefore we need to sum up
$O(1)$ quantities in this counting to all orders.

Such a resummation of the leading contributions to all orders
has been done by the following method.
The propagator for the conformal mode
can be read off from the action \eq{renac} for
small $\epsilon$ and $G$ as:
\beq
<\psi (p)\psi (-p)> = {G\over p^2}
e^{{\epsilon\over 2}\bar{\phi}},
\label{prop}
\eeq
where we have adopted the gauge fixing term which
eliminates the mixing between $h_{\mu\nu}$ and $\psi$
fields.
We have also shown the dependence on the
background conformal factor
($e^{-\bar{\phi}}$) explicitly.
The  divergent part of the vacuum expectation value of
the square of the $\psi$ field is:
\beq
<\psi ^2 > = {1\over \pi \epsilon} {G\over 2 }
e^{{\epsilon\over 2}\bar{\phi}} .
\label{div}
\eeq

We evaluate the anomalous dimension of this operator
by using the propagator \eq{prop} for the conformal mode.
In order to do so, we may utilize a zero dimensional model
with the following action which reproduces \eq{div}:
\beq
{a^2\over G}\pi\epsilon\psi ^2
e^{-{\epsilon\over 2}\bar{\phi}} .
\eeq
Then the vacuum expectation value of the cosmological constant
operator is:
\beq
<\sqrt{g}> = \int d\psi e^{-{D\over 2}\bar{\phi}}
exp({4\over \epsilon}log
(1+{\epsilon\over 4a}\psi ) -{1\over G}\pi\epsilon\psi ^2
e^{-{\epsilon\over 2}\bar{\phi}}) .
\eeq
This integral can be evaluated exactly for small $\epsilon$ by the
saddle point method after scaling the integration variable $\psi$
by $\epsilon$.

In this way the divergent part of the integral is found to be:
\beq
exp({4\over\epsilon}log(1+\rho _0) -
{16a^2\pi\over  G\epsilon}\rho _0^2
e^{-{\epsilon\over 2}\bar{\phi}}) ,
\eeq
where
\beq
\rho _0 = {1\over 2}(-1+\sqrt{1+{ G \over 2a^2\pi}}) .
\label{rho0}
\eeq
The anomalous dimension is found by
inspecting the $\bar{\phi}$ dependence
of this result to be
\beq
\gamma = \rho _0^2
{16a^2 \pi\over  G} = 2 - {16a^2 \pi\over  G}\rho _0 .
\eeq
By combining the canonical dimension of the
cosmological constant operator which is
the classical $\bar{\phi}$ dependence of it,
the scaling dimension
of this operator is found to be
${16a^2 \pi\over  G}\rho _0$
to the leading order in $\epsilon$.
At the short distance fixed point, it behaves as
$a\sqrt{{32\pi\over G}}$.

Since our results in this section
such as \rref{rho0}
involve the inverse power of $a^2$,
it is crucial to regard $a^2$ as an
independent and a finite coupling.
Although it is $O(\epsilon )$ classically,
it receives quantum corrections of $O(G)$.
In fact we have assumed that it is
as small as $G$ and resummed $O(1)$ quantities
in such a counting to all orders
by the saddle point method.

We now consider a physical $\beta$ function
of the theory.
One of the physical definition of the $\beta$ function
is to compare the gravitational coupling constant
$G$ and the cosmological constant $\Lambda$
\cite{KN,KKP}.
If we divide \eq{betan} by the scaling dimension of
the cosmological constant operator, we obtain
${aG^2\over \sqrt{8\pi G}}$. This is the $\beta$ function of the
gravitational coupling constant when we choose the
cosmological constant operator as the standard of the scale.

Although the cosmological constant operator
is multiplicatively renormalizable, we have found that
the quantum corrections are $O(1)$.
Hence it may be useful to define the renormalized cosmological
constant operator
including the quantum corrections which are as large as the
naive tree action.

The functional form of the renormalized cosmological constant
operator may be very generic.
However it can be fixed by the following method.
Here we adopt the strategy which has been successful
for two dimensional quantum gravity\cite{Polyakov,DDK}.
We decompose the metric into the background metric
$\hat{g} _{\mu\nu}$ and
the fluctuations around it.
In quantum gravity
physical observables should be background independent.
The functional form of the cosmological constant
operator
in terms of the conformal mode $\psi$ should be
such that it satisfies this requirement.

Let the renormalized cosmological constant
operator to be
\beq
\int e^{-{D\over 2}\bar{\phi}}\Lambda (\psi ).
\label{rencos}
\eeq
We assume that the renormalized
cosmological constant operator is invariant under the volume
preserving diffeomorphism just like the tree action.
It is also reasonable to assume that it only depends
on $\psi$ and $\sqrt{\hat{g}} = e^{-{D\over 2}\bar{\phi}}$.
We expect that the counter terms may depend on generic fields.
We parametrize the cosmological constant operator as:
\beq
\Lambda (\psi ) = exp(\alpha \psi + {1\over 2} \beta \psi ^2
\ldots ).
\eeq
In order to determine $\Lambda (\psi )$,
we impose the invariance under the gauge transformation
\eq{gaugetr} on this operator.
If the theory is invariant under the volume preserving
diffeomorphism,
the gauge transformation \rref{gaugetr}
holds if the theory is invariant under the
the conformal transformation \rref{conftr}.
For this reason,
we only need to require
the invariance under the conformal
transformation \eq{conftr}
in order to impose the gauge invariance \rref{gaugetr}.


The important quantum effect is the anomalous dimension
of the operator $\Lambda (\psi )$.
At the one loop order, $\Lambda (\psi )$
changes under the transformation \eq{conftr} as:
\beq
\delta \Lambda (\psi ) = {G\over 8\pi} {\del ^2\over
\del \psi ^2}\Lambda (\psi ) \delta \bar{\phi} .
\eeq
The variation due to the $\psi$ field transformation as in
\eq{conftr} is
\beq
\delta \Lambda (\psi ) = (D-1)(a+2\epsilon b \psi)
{\del \over \del \psi} \Lambda (\psi ) \delta \bar{\phi} .
\eeq
The sum of the above must cancel the variation
of $\delta (\sqrt{\hat{g}})\Lambda = -{D\over 2}\delta \bar{\phi}
(\sqrt{\hat{g}}) \Lambda $.
The coefficients $\alpha ,\beta ,\ldots $ are determined
in this way as:
\beq
\alpha ={4\pi a\over G} (-1\pm\sqrt{1+{G\over 2\pi a^2}}),
\beta = -{\epsilon \alpha \over 4a + G\alpha/\pi},
\ldots
\label{albe}
\eeq
We choose the $+$ sign out of the two possible
branches in the above expression since it
possesses the correct semiclassical limit.

{}From the tree action \eq{renac}, we read off the effective
gravitational coupling as the multiplication factor
in front of $\tilde{R}$:
\beq
{1\over G} -{\delta G\over G^2}
= {1\over G}(1 + a\psi + \epsilon b \psi ^2) .
\eeq
On the other hand, the log of the cosmological constant operator
is found to be:
\beq
log(\Lambda ) = \alpha \psi + {1\over 2}\beta \psi ^2 \ldots
\eeq
Since the quantum fluctuation of $\psi$ is $O(\sqrt{G})$,
it is at most $O(\sqrt{\epsilon})$
even around the short distance fixed point.
There is a scaling window for $\psi < 1/\sqrt{\epsilon}$
since the nonlinear terms of $\psi$ in the above
expressions can be neglected then.
{}From these reasonings, the $\beta$ function is found to be
\beq
{d G\over d log(\Lambda ) }= {aG\over \alpha} .
\eeq
At the short distance fixed point,
$\alpha \rightarrow \sqrt{8\pi /G}$.
Hence the above $\beta$ function
approaches
$aG^2/\sqrt{8\pi G}$
which agrees with our result in the first part of this
section.

The relationship between the bare cosmological constant
operator \rref{barecos} and the renormalized one
\rref{rencos} may be understood as follows.
We conjecture that the bare cosmological constant operator
which is obtained after adding all necessary counter terms to the
renormalized operator is manifestly invariant under the diffeomorphism.
The precise relationship at the one loop level is:
\beq
\int d^Dx \sqrt{g}  = \int exp(-{G\over 4\pi \epsilon}
{\del ^2\over \del \psi ^2})\Lambda (\psi ) .
\label{poscos}
\eeq
{}From this equation the renormalization group
equation for $\Lambda (\psi )$ follows:
\beq
{\mu {\del\over \del \mu}}\Lambda (\psi )
={G\over 4\pi } {\del ^2\over \del \psi ^2}
\Lambda (\psi ) ,
\eeq
where we have used the leading renormalization
group equation
$\mu{d\over d\mu}G = \epsilon G$.

The solution of this diffusion equation is
\beq
\Lambda (\psi ) = \int_{-\infty}^{\infty} d\psi '
\sqrt{{\epsilon \over G}}
exp(-{\pi \epsilon\over G}(\psi -\psi ')^2)
\Lambda _I(\psi ') ,
\eeq
where $\Lambda _I(\psi )$ is the initial condition
of the renormalized operator.
For the initial condition, we can assume the classical
expression which is the same with the bare expression
\rref{barecos} at the weak coupling limit.
This is because there should be no renormalization
when the coupling is very weak. It naturally follows
from our postulate \rref{poscos}.

We can evaluate this integral by the saddle point
method again for small $\epsilon$ limit:
\beqa
\Lambda (\psi ) & = &
\int _{-\infty}^{\infty} d \rho
exp( -{16\pi a^2\over \epsilon G}(\rho -{\epsilon\over 4a}
\psi )^2 + {4\over \epsilon}log(1+\rho )) \n
& \sim &
exp( \alpha \psi -{1\over 2 }{\epsilon\pi\over G}\psi ^2 ) ,
\eeqa
where $\alpha$ is precisely the same coefficient
with \rref{albe}. The coefficient of $\psi ^2$ term
is estimated
at short distance where $a$ is small which
also agrees with \rref{albe}.
Therefore we have derived the functional form of the
renormalized cosmological constant operator from the
bare cosmological constant operator based on the
postulate \rref{poscos} at the one loop level.

The $\beta$ function of $G$ with
respect to the cosmological constant
we have studied in this section differs from the
$\beta$ function \eq{betan} in the previous section
which is related to the conformal anomaly.
This situation reflects the inherent ambiguity of the definition
of the $\beta$ functions in quantum gravity.
Namely they differ when different operator is chosen as the
standard of the scale.
However the both quantities vanish at the short distance fixed
point where $a=0$. From this sense the both definitions are
consistent.

\section{Conclusions and Discussions}

In this paper we have further studied the
renormalizability of quantum gravity near two
dimensions.
We thereby put the $2+\epsilon$
dimensional expansion of quantum gravity
on a solid foundation.
We have proven that all necessary counter terms can be
supplied by the bare action which is invariant
under the full diffeomorphism.

However the tree level action itself is not
invariant under the general coordinate transformation.
Only after adding the counter terms and thereby considering
the bare action, we can recover the action which is
invariant under the full diffeomorphism.

We have chosen the tree level action to possess the
volume preserving diffeomorphism invariance.
In order to recover the full diffeomorphism invariance,
we need to require that the theory is independent
of the background metric.
This requirement has led us to search a theory which
is conformally invariant with respect to the background metric.
Obviously the Einstein action is such a theory and
we conjecture that the requirement of the background independence
leads us uniquely to the Einstein action as the bare action.

In our perturbative expansion, we need to introduce
not only the gravitational coupling constant $G$ but also
another coupling constant $a$.
$G$ controls the dynamics of $h_{\mu\nu}$ field
while $a$ controls the dynamics
of the conformal mode. $a^2$ is related to $G$n
since it is nothing but the $\beta$ function of $G$
modulo a factor of $G$.
However
$a$ itself cannot be expanded in $G$ since
such an expansion is singular in $\epsilon$.

We have constructed a proof of the renormalizability
of the theory to all orders in the perturbative
expansion of $G$. In this expansion, $a^2$ is also
perturbatively determined in terms of $G$.
Since $G$ is $O(\epsilon )$ at the short distance
fixed point of the renormalization group,
$G$ may be regarded as a small expansion parameter
as long as we regard $\epsilon$ to be small.
Our proof is base on the plausible assumption
concerning the solution of \eq{s*g}.
This assumption is in accord with the investigations of the
gauge theories\cite{ZJ,BH}. We hope that it can also be
proven in the near future.

The renormalization of the cosmological
constant operator is analogous.
We have shown that the cosmological constant operator
is multiplicatively renormalizable.
The anomalous dimension of the cosmological
constant operator is generically  $O(1)$ near two dimensions
and we need to be more careful to calculate it.
However it can be calculated for small $\epsilon$ by the
saddle point method. The exact two dimensional solutions
can be understood in this way\cite{KKN1}.
It may be useful to consider the renormalized cosmological constant
operator which incorporates the quantum effect.
We can require the background metric
independence to determine the renormalized
cosmological constant operator
in the analogous way
with the two
dimensional quantum gravity.
The relationship between the bare and the renormalized
cosmological constant operator is also considered.
We need to add counter terms to the renormalized
cosmological constant operator.
We have postulated that the bare cosmological constant operator
which is obtained in this way
is manifestly invariant under the diffeomorphism.

In order to calculate the $O(\epsilon )$ corrections,
we need to perform the full two loop calculations.
The partial results have been reported in \cite{AKNT}.
We believe that we have made it clear how such a
calculation can be completed.

\vspace{10mm}

This work is supported in part by the Grant-in-Aid for Scientific Research
from the Ministry of Education, Science and Culture.
One of us (Y.K.) acknowledges the Aspen Center for Physics
where a part of this work has been carried out.

\newpage
\setlength{\baselineskip}{7mm}

\end{document}